\documentclass[]{spie}  

\usepackage{amsmath,amsfonts,amssymb}
\usepackage{graphicx}
\usepackage[colorlinks=true, allcolors=blue]{hyperref}

\title{Secret key capacity of the thermal-loss channel: \\Improving the lower
bound}

\author{Carlo Ottaviani, Riccardo Laurenza, Thomas P. W. Cope, Gaetana Spedalieri,\\ Samuel L. Braunstein, Stefano
Pirandola} \affil{Computer Science and York Centre for Quantum
Technologies, \\University of York, York YO10 5GH, United Kingdom}
\begin{document}
\maketitle

\begin{abstract}
We consider the secret key capacity of the thermal loss channel,
which is modeled by a beam splitter mixing an input signal mode
with an environmental thermal mode. This capacity is the maximum
value of secret bits that two remote parties can generate by means
of the most general adaptive protocols assisted by unlimited and
two-way classical communication. To date, only upper and lower
bounds are known. The present work improves the lower bound by
resorting to Gaussian protocols based on suitable trusted-noise
detectors.
\end{abstract}



\section{Introduction and relations with previous works}

In recent years, the field of quantum information and
communication has witnessed a growing interest in the area of
continuous variable (CV) quantum systems~\cite{BraRMP}.
Particularly successful has been the development of Gaussian
quantum information~\cite{RMP} and the area of Gaussian quantum
key distribution (QKD)~\cite{LeverrierEntropy}. The latter focuses
on the implementation of protocols exploiting Gaussian states and
operations. Gaussian states are easy to study because they are
just characterized by their first two statistical moments, i.e.,
mean value and covariance matrix (CM)~\cite{RMP}. Furthermore,
they are the easiest states that can be generated in quantum
optics labs.

The research in Gaussian QKD has led to the design, the
theoretical analysis and the experimental implementation of secure
protocols in
one-way~\cite{GrosshansNAT,Chris1,Chris2,Lutkenaus,PC-PRL2009,Thermal1,Thermal2,Thermal3,Thermal4,Thermal5,Thermal6,Usenko,prova1,prova2,prova3},
two-way~\cite{pirs2way,Ott2way1,Ott2way2,Thermal2way} and, more
recently, measurement-device-independent
configurations~\cite{CVMDIQKD,PRA-S}. Today state-of-the-art
implementations of Gaussian QKD have been proven to be secure over
distances of about $80$ km~\cite{diamanti}. More so than distance,
high rates are the promise of Gaussian protocols, with potential
performances which are orders of magnitude higher than their
discrete-variable (DV) counterpart~\cite{CVMDIQKD-reply}. High
rates are important not only for secure quantum networks but, more
generally, for any realistic design of a quantum
Internet~\cite{Kimble,StefanoSamNature}.

From a theoretical point of view, it is therefore of fundamental
importance to determine the optimal secret key rate that is
achievable by two remote parties, say Alice and Bob, at the two
ends of a quantum channel, especially if this channel is Gaussian.
This optimal rate (capacity) is achieved by optimizing over the
most general QKD protocols, which are based on local operations
(LOs) assisted by unlimited two-way classical communication (CC),
briefly called adaptive LOCCs. The presence of feedback makes
these protocols very hard to study but, recently,
Ref.~\citen{stretching} has introduced a novel methodology which
fully simplifies the analysis of such protocols for the most
important quantum channels for both CV and DV systems.

In fact, Ref.~\citen{stretching} has brought two important
advances: (i) it has extended the notion of relative entropy of
entanglement (REE) from quantum states to quantum channels; (ii)
it has devised a novel technique based on
teleportation~\cite{Tele1,Tele2,AdvTel}, called `teleportation
stretching', able to reduce an arbitrary adaptive protocol into a
much simpler block form, as long as the protocol is implemented at
the ends of a quantum channel which suitably commutes with
teleportation. By combining these two ingredients,
Ref.~\citen{stretching} has computed the tightest-known upper
bounds for the two-way quantum capacity $Q_2$ and the secret key
capacity $K$ of many quantum channels, including DV Pauli
channels, DV erasure channels and CV bosonic Gaussian channels. In
particular, by showing coincidence with suitable lower bounds,
Ref.~\citen{stretching} has established the secret-key capacities
of dephasing channels, erasure channels, quantum-limited
amplifiers and lossy channels. (For the erasure channel see also
the independent derivation of Ref.~\citen{GEW} based on the
different tool of the squashed
entanglement~\cite{squashed,squashed2,squashed3}.)

In the specific case of a lossy channel with arbitrary
transmissivity $\eta$, Ref~\citen{stretching} has derived
\begin{equation}
Q_2(\eta)=K(\eta)=-\textrm{log}_2(1-\eta).
\end{equation}
For high loss (i.e., long distances), this provides about $1.44$
secret bits per use, which is the fundamental rate-loss scaling
which restricts any secure quantum optical communications. This is
also the benchmark that a quantum repeater must surpass in order
to be a meaningful device. The generalization of this scaling to
repeater-assisted quantum communications and multi-hop quantum
networks has been recently provided by Ref.~\citen{optimalNET}.
(See also Ref.~\citen{stretching3} for the specific study of
single-hop quantum communication networks.)

An important bosonic Gaussian channel which has not been fully
solved by Ref~\citen{stretching} is the thermal-loss channel,
which can be modeled by a beam splitter with transmissivity
$\eta$, mixing the input signal mode with an environmental mode
described by a thermal state with variance $\omega=2
\bar{n}_{th}+1$, with $\bar{n}_{th}$ being the mean number of
thermal photons. W still do not know the two-way quantum capacity
$Q_2(\eta,\omega)$ and the secret key capacity $K(\eta,\omega)$ of
this channel. We only know an upper bound, provided by the
REE+teleportation method of Ref.~\citen{stretching}, and a lower
bound, first computed by Ref.~\citen{reversePRL} by using the
notion of reverse coherent information~\cite{reverseCoh}. In
particular, these bounds provide the following sandwich for the
secret key capacity
\begin{align}
-\log_{2} \left( 1-\eta\right) -h(\omega)\leq K(\eta,\omega)\leq
\Phi(\eta,\omega)~, \label{LBUB}
\end{align}
where \begin{equation}
h(x):=\frac{x+1}{2}\log_{2}\frac{x+1}{2}-\frac{x-1}{2}\log_{2}\frac{x-1}{2},
\label{h-DEF}
\end{equation}
and
\begin{align}  \label{UB}
\Phi(\eta,\omega):=\left\{
\begin{array}{rl}
& -\log_2[(1-\eta)\eta^{\bar n_{th}}]-h(\omega)\quad\text{for}~\bar n_{th}<%
\frac{\eta}{1-\eta} \\
&  \\
& 0\qquad\text{otherwise}%
\end{array}
\right.
\end{align}

In this paper, we improve the lower bound present in
Eq.~(\ref{LBUB}). Our improved bound applies to the secret key
capacity only (and not to the two-way quantum capacity), because
it is based on the optimization of a class of QKD protocols. In
fact, we consider a type of noise-assisted Gaussian QKD protocol
which is a coherent version of the one considered in
Ref.~\citen{PC-PRL2009} (because it involves a quantum memory) and
a direct generalization of the one studied in
Ref.~\citen{reversePRL} (because it considers more general noise
at the detection stage). In this protocol, Alice distributes one
mode of a two-mode squeezed vacuum (TMSV) state through the
thermal-loss channel, while keeping the other mode in a quantum
memory. This is repeated many times. At the output of the channel,
Bob's measurement setup consists of a beam splitter with arbitrary
transmissivity and subject to a trusted-noise thermal environment;
this is followed by homodyne detection, randomly chosen in one of
the two quadratures. At the end of the quantum communication, Bob
communicates which quadrature he has measured in each round. As a
consequence, Alice performs the same sequence of homodyne
detections on her systems and tries to infer Bob outcomes (reverse
reconciliation). We compute the rate of this protocol, and we
consider its optimization over the free parameters at Bob's side.
This will provide our improved lower bound.

The paper is structured as follows. In Sec.~\ref{adPro}, we review
the definition of the secret key capacity of a quantum channel. In
Sec.~\ref{CH2}, we review the main techniques and results of
Ref.~\citen{stretching}, which have led to the lower and upper
bounds for the secret key capacity of the thermal-loss channel. In
Sec.~\ref{scheme}, we show the result of this work, i.e., how to
improve the lower bound by means of the noise- and memory-assisted
Gaussian QKD protocol. Finally, Sec.~\ref{CONCLUSIONS} is for the
conclusions.


\section{Secret key capacity of a quantum channel}\label{adPro}

Assume that Alice and Bob are connected by a quantum channel
$\mathcal{E}$ over which they perform an adaptive protocol (see
Ref.~\citen{stretching} for full details). After $n$ transmissions
through $\mathcal{E}$, the two parties share an output state
$\rho_{\mathbf{ab}}^n:=\rho_{\mathbf{ab}}(\mathcal{E}^{\otimes
n})$  which depends on the sequence $\mathcal{L}=\{\Lambda_0,%
\Lambda_1,\ldots,\Lambda_n\}$ of adaptive LOCCs performed. The protocol is characterized by the triplet $(n,\epsilon,R^n)$ where  $%
\epsilon$ and the rate of the protocol $R^n$ are such that  $||\rho_{\mathbf{%
ab}}^n-\phi_n||\leq\epsilon$  where $\phi_n$ is a target state
with a content of information equal to $nR^n$ bits. By optimizing
over all the possible LOCC-sequences $\mathcal{L}$ and by taking
the limit of infinite channel uses $n\rightarrow\infty$, one
defines the generic two-way capacity of the channel as follows
\begin{align}  \label{genericC}
\mathcal{C}(\mathcal{E}):=\sup_\mathcal L\lim_{n\rightarrow\infty}R^n~.
\end{align}

Here we are interested in the optimal performance that can be
achieved by QKD protocols over the quantum channel $\mathcal{E}$.
In this scenario the two parties aim at generating secret bits, so
that the target state $\phi_n$ is a private state~\cite{private}.
The generic two-way capacity of Eq.~(\ref{genericC}) automatically
defines the secret key capacity
$\mathcal{C}(\mathcal{E})=K(\mathcal{E})$ of the channel: This is
the maximum achievable number of secret bits that can be
transmitted per channel use. In general the computation of the
two-way capacities defined in Eq.~(\ref{genericC}) is extremely
demanding. This is mainly due to the fact that we have to optimize
over a family of LOCCs which entails feedback, which are exploited
to optimize the subsequent inputs to the channel.

\section{Bounds for the secret key capacity}\label{CH2}

\subsection{Lower bound}

The best-known lower bounds for the secret key capacity
$K(\mathcal{E})$ are given by the (reverse) coherent information
$I_{(RC)C}(\mathcal{E})$ of the channel. Consider a maximally
entangled state of systems $A$ and $B$, i.e. an EPR state
$\Phi_{AB}$. The propagation of half of such a state through the
channel $\mathcal{E}$ defines its Choi matrix~\cite{Choi}
$\rho_{\mathcal{E}}:=(\mathcal{I}\otimes\mathcal{E})\Phi_{AB}$.
This allows us to introduce the coherent
information~\cite{Coh1,Coh2} of the channel $I_C(%
\mathcal{E})$ and its reverse
counterpart~\cite{reverseCoh,reversePRL} $I_{RC}(\mathcal{E})$
which are defined as
\begin{align}  \label{RevCo}
I_{C(RC)}(\mathcal{E}):=S[\text{Tr}_{A(B)}(\rho_\mathcal E)]-S(\rho_\mathcal
E)~,
\end{align}
where $S$ is the von Neumann entropy. As a consequence of the
hashing inequality~\cite{hash}, one may write (see
Ref.~\citen{stretching} for more details)
\begin{align}
\max\{I_C(\mathcal{E}),I_{RC}(\mathcal{E})\}\leq K(\mathcal{E})~.
\end{align}


For the specific case of the thermal-loss channel, the best lower
bound is that provided by the reverse coherent information. In a
thermal-loss channel $\mathcal{E}(\eta,\omega)$, the input signals
are combined with thermal noise $\omega=2\bar{n}_{th}+1$, i.e.,
the input quadratures are transformed according to
\begin{align}
\hat x\rightarrow\sqrt{\eta}\hat x+\sqrt{1-\eta}\hat x_E~,
\end{align}
where $0\leq\eta\leq 1$ is the transmissivity and $\hat x_E$ is
the environment in a thermal state with $\bar{n}_{th}$ mean photon
number. The Choi matrix of this bosonic channel is
energy-unbounded, so that it should be intended as an asymptotic
limit of a suitable sequence of finite-energy states. In
particular,
the CV EPR asymptotic state $\Phi$ is defined as the limit for $%
\mu\rightarrow\infty$ of two-mode squeezed vacuum (TMSV) states~\cite{RMP} $\Phi^\mu=|\Phi^\mu\rangle\langle\Phi^%
\mu|$. In other words, we have
\begin{equation} \label{limitChoi}
\rho_{\mathcal{E}}:=\lim_{\mu\rightarrow\infty}\rho^\mu~,~~~
 \rho^\mu=(\mathcal{I}\otimes\mathcal{E})\Phi^\mu~~.
\end{equation}

As a consequence, the reverse coherent coherent information of
Eq.~(\ref{RevCo}) is computed as follows
\begin{equation}
I_{RC}(\mathcal{E})=\lim_{\mu\rightarrow\infty}I_{RC}(\mathcal{E},\rho^\mu)=\lim_{\mu\rightarrow\infty}S[\text{Tr}_{B}(\rho^\mu)]-S(\rho^\mu)~.
\end{equation}
A direct evaluation for the thermal-loss channel leads to the
following expression~\cite{reversePRL,stretching}
\begin{equation}  \label{LB}
I_{RC}(\eta,\omega)=-\log_{2} \left( 1-\eta\right) -h(\omega).
\end{equation}



\subsection{Upper bound\label{Upper}}

As previously mentioned, a general upper bound for the secret key
capacity of a quantum channel can be designed by extending the
notion of relative entropy of entanglement (REE) from quantum
states to quantum channels. Recall that for any bipartite state
$\rho$ the REE is defined as $E_R(\rho)=\min_\sigma
S(\rho||\sigma)$, where the minimization is taken over the set of
all the
possible separable states and $S(\rho||\sigma):=\text{Tr}[\rho(\log_2\rho-%
\log_2\sigma)]$ is the relative entropy.\newline By exploiting the
properties of the REE, Ref.~\citen{stretching} showed that, at any
dimension, the generic two-way capacity of Eq.~(\ref{genericC}) is
upper bounded as follows
\begin{align}  \label{adaUB}
\mathcal{C}(\mathcal{E})\leq E_R^\star(\mathcal{E}):=\sup_\mathcal
L\limsup_{n\rightarrow\infty}\frac{E_R(\rho_{\mathbf{ab}}^n)}{n},
\end{align}
where $E_R^\star(\mathcal{E})$ is called the adaptive REE of the
channel. The latter is very hard to compute. However, for a wide
class of quantum channels, called teleportation-covariant or
`stretchable' channels~\cite{stretching}, the adaptive REE of the
channel $\mathcal{E}$ is simply bounded by the REE of its Choi
matrix, i.e., one may write~\cite{stretching}
\begin{align} \label{crucial}
E_R^\star(\mathcal{E})\leq E_R(\rho_\mathcal E)~,
\end{align}
so that the computation of the upper bound in Eq.~(\ref{adaUB}) is
reduced to the much simpler computation of the single-letter
quantity $E_R(\rho_\mathcal E)$. Because this quantifies the
amount of entanglement (REE) distributed through the channel by an
EPR state, it can be called the \textit{entanglement flux} of the
channel~\cite{stretching} and is denoted as follows
\begin{align}
\Phi(\mathcal{E}):=E_R(\rho_\mathcal E)~.
\end{align}
Let us provide more details behind the crucial simplification of
Eq.~(\ref{crucial}) which relies on the technique of teleportation
stretching devised in Ref.~\citen{stretching}.

\subsubsection{Teleportation simulation of a channel and teleportation stretching of an adaptive protocol\label{Stretch}}

Teleportation stretching allows us to reduce an adaptive protocol
with an arbitrary associated quantum task (quantum information
transmission, entanglement distribution or QKD) into an equivalent
non-adaptive protocol, performing exactly the same original task
but whose output state is expressed in a convenient block form.
This reduction process can be exploited at any dimension, finite
or infinite, whenever the quantum channel suitably commutes with
the set $\mathbb{U}_d$ of the teleportation unitaries in dimension
$d$. At finite dimension, the elements of $\mathbb{U}_d$ are
represented by generalized Pauli
operators. For CV systems $%
(d\rightarrow\infty)$, the set $\mathbb{U}_\infty$ is composed of
displacement operators~\cite{AdvTel}.

By definition, a quantum channel $\mathcal{E}$ is called
``teleportation-covariant'' or simply ``stretchable'' if, for any
teleportation unitary $U\in\mathbb{U}_d$, we can
write~\cite{stretching}
\begin{align}
\mathcal{E}(U\rho U^\dagger)=V\mathcal{E}(\rho)V^\dagger
\end{align}
for some unitary $V$. The key property of stretchable channels is
represented by the fact that the teleportation unitaries can be
pushed out of the channel and mapped into generally-different
unitaries that can still be corrected. As a consequence of this
property, the transmission of a quantum state through the channel
$\mathcal{E}$ can be simulated by teleporting the state using the
Choi matrix of the channel $\rho_\mathcal E$.
Ref.~\citen{stretching} developed this teleportation-simulation
argument for both DV and CV channels. (Note that the teleportation
simulation of the specific class of DV Pauli channels was
previously discussed in Refs.~\citen{Bennett96,Bose01}).

Teleportation-simulation is only the first step of teleportation
stretching. The technique of teleportation stretching involves the
following main steps~\cite{stretching}: (i) first the stretchable
channel is replaced with teleportation over its Choi matrix; (ii)
teleportation is considered as an additional LOCC, while the Choi
matrix is anticipated back in time and stretched out of the
adaptive LOCCs; (iii) all the adaptive LOCCs are collapsed into a
single trace preserving LOCC. By applying this procedure
iteratively for all the $n$ transmissions, the output state shared
by Alice and Bob can be expressed as follows~\cite{stretching}
\begin{align} \label{Choidec}
\rho_{\mathbf{ab}}^n=\bar\Lambda(\rho^{\otimes n}_\mathcal E)
\end{align}
in terms of a single and very complicated trace preserving LOCC
$\bar\Lambda$. (Note that this result must not be confused with
the content of Section V of Ref.~\citen{Bennett96} which instead
showed how to transform a quantum communication protocol into an
entanglement distillation protocol).

Let us now combine the Choi decomposition of Eq.~(\ref{Choidec})
with Eq.~(\ref{adaUB}). Using the properties of the REE, one
derives~\cite{stretching}
\begin{align}  \label{redu}
E_R(\rho_{\mathbf{ab}}^n)\overset{(1)}{\leq}E_R(\rho^{\otimes n}_\mathcal E)%
\overset{(2)}{\leq}nE_R(\rho_\mathcal E),
\end{align}
where $(1)$ the REE is monotonically decreasing under trace
preserving LOCCs and $(2)$ it is subadditive over tensor products.
Thus, the complicated trace preserving LOCC $\bar\Lambda$
disappears in the chain of Eq.~(\ref{redu}). Then, by substituting
Eq.~(\ref{redu}) into Eq.~(\ref{adaUB}), one can also simplify the
the upper limit and the supremum in the definition of
$E_R^\star(\mathcal{E})$, obtaining~\cite{stretching}
\begin{align}
\mathcal{C}(\mathcal{E})\leq E_R^\star(\mathcal{E})\leq
E_R(\rho_{\mathcal{E}}):=\Phi(\mathcal{E})~,
\end{align}
Thus, the secret key capacity of a stretchable channel is upper
bounded by its entanglement flux, which is a single-letter
computable quantity. This final result is achieved by combining
the upper bound given by the adaptive REE of the channel with the
technique of teleportation stretching.

\subsubsection{Entanglement flux of the thermal-loss channel\label{flux}}

It is easy to verify that bosonic Gaussian channels are
stretchable. For a Gaussian channel, we can formally define its
entanglement flux $\Phi(\mathcal{E})=E_R(\rho_\mathcal E)=\min_{\sigma \in\text{SEP}%
}S(\rho_\mathcal E||\sigma)$ as~\cite{stretching}
\begin{align}
\Phi(\mathcal{E})&:=\min_{\{\sigma^\mu\}}\liminf_{\mu\rightarrow\infty}S(\rho^\mu||\sigma^\mu),
\end{align}
where $\rho^{\mu}$ is given in Eq.~(\ref{limitChoi}) and the
minimization is over all sequences $\{\sigma^\mu\}$ of separable
states that converge in trace norm. This formulation takes into
account of the fact that the Choi matrix of a Gaussian channel is
energy unbounded. Therefore, the secret key capacity of a Gaussian
channel satisfies~\cite{stretching}
\begin{align}
K(\mathcal{E})\leq E^\star_R(\mathcal{E})\leq\Phi(\mathcal{E}%
)\leq\liminf_{\mu\rightarrow\infty}S(\rho^\mu||\widetilde\sigma^\mu)
\end{align}
for some (good) sequence of separable Gaussian states
$\{\widetilde\sigma^\mu\}$. This choice is indeed crucial for
having a good upper bound. By exploiting a simple formula for the
relative entropy between Gaussian states proven in
Ref.~\citen{stretching}, one may compute
$S(\rho^\mu||\widetilde\sigma^\mu)$. For the thermal-loss channel,
one finds the upper bound
\begin{equation}
\mathcal{C}(\mathcal{E})\leq \Phi({\eta,\omega}),
\end{equation}
where $\Phi({\eta,\omega})$ is given in Eq.~(\ref{UB}).


\section{New lower bound for the secret key capacity of the thermal-loss channel}\label{scheme}

In this paper, we show the following result.
\begin{theorem} Consider a thermal-loss channel with transmissivity $\eta$ and thermal noise
$\omega$. Its secret key rate $K(\eta,\omega)$  is lower-bounded
by
\begin{equation}
R_{M}(\eta,\omega)=\underset{\eta_{d},\gamma}{\max}~R\left(
\eta,\omega,\eta_{d},\gamma\right) ,   \label{RM}
\end{equation}
where
\begin{align}
R\left( \eta,\omega,\eta_{d},\gamma\right) &:=\frac{1}{2}\log_{2}
\frac{\eta_{d}\omega+(1-\eta_{d})(1-\eta)\gamma}{\left(
1-\eta\right) \left[ \eta_{d}(1-\eta)\omega+\left(
1-\eta_{d}\right) \gamma\right] }+ h\left(
\sqrt{\frac{\omega\left[ \eta_{d}+(1-\eta)\left( 1-\eta_{d}\right)
\omega\gamma\right] }{\eta_{d}\omega+(1-\eta)\left(
1-\eta_{d}\right) \gamma}}  \right) -h(\omega), \label{rate}
\end{align}
and the maximization is over detection transmissivity $\eta_{d}$
and thermal variance $\gamma\geq1$.
\end{theorem}

\bigskip

\begin{figure}[th]
\vspace{-1.5cm}
\par
\begin{center}
\includegraphics[height=7.5cm]{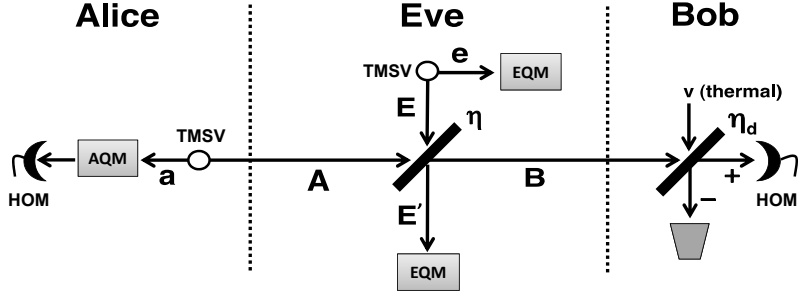}
\end{center}
\par
\vspace{-1.5cm} \caption[example]{Noise-assisted Gaussian
protocol.  HOM = homodyne detection, TMSV = two-mode squeezed
vacuum state, EQM = Eve's quantum memory, AQM = Alice's quantum
memory. See text for explanations.} \label{Scheme-Fig1}
\end{figure}

\textbf{Proof.~~}Consider the Gaussian protocol described in
Fig.~\ref{Scheme-Fig1}. Alice has a TMSV state $\Phi ^{\mu }$ of modes $a$ and $A$%
. This is a zero mean Gaussian state with CM%
\begin{equation}
\mathbf{V}_{aA}=\left(
\begin{array}{cc}
\mu \mathbf{I} & \sqrt{\mu ^{2}-1}\mathbf{Z} \\
\sqrt{\mu ^{2}-1}\mathbf{Z} & \mu \mathbf{I}%
\end{array}%
\right) :=\mathbf{V}_{\mathrm{TMSV}}(\mu ),  \label{Valice}
\end{equation}%
where $\mathbf{I=}$\textrm{diag}$(1,1)$ and
$\mathbf{Z=}$\textrm{diag}$(1,-1) $. Mode $a$ is kept in a quantum
memory (for later measurement). Mode $A$ is sent through the
thermal-loss channel with transmissivity $0\leq \eta \leq 1$ and
thermal noise $\omega =2\bar{n}_{th}+1$. At the output, Bob
implements a noisy detection. He uses a beam splitter of
transmissivity $\eta _{d}$ to mix the output mode $B$ with a mode
$v$ in a thermal state with variance $\gamma $, i.e., with CM
\begin{equation}
\mathbf{V}_{v}=\gamma \mathbf{I}.  \label{Vth}
\end{equation}%
The output $+$ of the beam splitter is then homodyned in the $q$
or $p$ quadrature, randomly. After (large) $n$ rounds, Bob
communicates which quadrature he has measured in each round, so
that Alice can perform exactly the same sequence of homodyne
detections on the $a$-modes she has kept. Her outcomes are finally
used to infer Bob's outcomes (reverse reconciliation).

In the middle, between the two parties, the thermal-loss channel
can be dilated into a beam splitter with transmissivity $\eta $
mixing the input mode $A$ with an environmental mode $E$, which is
described by a thermal state with variance $\omega $. This thermal
state can then be purified into a TMSV state of modes $e$ and $E$,
which is completely under Eve's control. This is
known as an entangling cloner~\cite{cloner}. The CM of Eve's input state is $\mathbf{V}%
_{eE}=\mathbf{V}_{\mathrm{TMSV}}(\omega )$. The output modes $e$ and $%
E^{\prime }$ are stored in a quantum memory which is coherently
measured by Eve at the end of the protocol (collective attack).

The initial global state of Alice, Bob and Eve is given by the
tensor
product $\rho _{0}=\rho _{aA}\otimes \rho _{eE}\otimes \rho _{v}$, with CM $%
\mathbf{V}_{0}=\mathbf{V}_{aA}\mathbf{\oplus V}_{eE}\mathbf{\oplus V}_{v}=%
\mathbf{V}_{aAeEv}$. For our convenience, we rearrange the modes
so to obtain $\mathbf{V}_{0}=\mathbf{V}_{aeEAv}$. This state is
processed by a sequence of two beam-splitters ($\eta $ and $\eta
_{d}$). First we process
mode $E$ and $A$, by applying the symplectic transformation $\mathbf{\tilde{V%
}}=\mathbf{S}_{\eta }\mathbf{V}_{0}\mathbf{S}_{\eta }^{T}$, where $\mathbf{S}%
_{\eta }:=\mathbf{I\oplus I\oplus T}(\eta )\oplus \mathbf{I}$,
with
\begin{equation}
\mathbf{T}(\eta ):=\left(
\begin{array}{cc}
\sqrt{\eta }\mathbf{I} & \sqrt{1-\eta }\mathbf{I} \\
-\sqrt{1-\eta }\mathbf{I} & \sqrt{\eta }\mathbf{I}%
\end{array}%
\right) .  \label{T}
\end{equation}%
Then, we compute%
\begin{equation}
\mathbf{V}=\mathbf{S}_{\eta _{d}}\mathbf{\tilde{V}S}_{\eta _{d}}^{T},~~%
\mathbf{S}_{\eta _{d}}:=\mathbf{I\oplus I\oplus I\oplus T}(\eta
_{d})~. \label{V-OUT}
\end{equation}%
The CM of Eq.~(\ref{V-OUT}) describes the global output state
$\rho _{aeE^{\prime }+-}$. Discarding mode $-$ corresponds to
considering the reduced state $\rho _{aeE^{\prime
}+}=\mathrm{Tr}_{-}(\rho _{aeE^{\prime }+-})$ with CM
$\mathbf{V}_{aeE^{\prime }+}$. From this CM\ we may compute Alice
and Bob's mutual information $I_{AB}$ as well as Eve's Holevo
information $\chi _{EB}$ on Bob's outcomes (reverse
reconciliation). Under ideal conditions of perfect reconciliation
efficiency, the key rate is $R=I_{AB}-\chi _{EB}$.

To compute $I_{AB}$ we select from $\mathbf{V}_{aeE^{\prime }+}$,
the block
relative to mode $+$ measured by Bob. This is given by $\mathbf{V}_{+}=V_{B}%
\mathbf{I}$, where
\begin{equation}
V_{B}=\eta _{d}\left[ \eta \mu +\left( 1-\eta \right) \omega
\right] +\left( 1-\eta _{d}\right) \gamma :=V(\mu ).  \label{VBOB}
\end{equation}%
Note that Alice's homodyne detection on mode $a$ of the TMSV state
(after it is released by the memory) is equivalent to preparing
Gaussianly-modulated squeezed states on mode $A$. Thus Bob's
conditional variance $V_{B|A}$ can be computed by simply setting
$1/\mu $ in Eq.~(\ref{VBOB}), i.e., we may write $V_{B|A}=V(1/\mu
)$. Thus, we get
\begin{equation}
I_{AB}=\frac{1}{2}\log \frac{V_{B}}{V_{B|A}}\rightarrow
\frac{1}{2}\log \frac{\eta _{d}\eta \mu }{\eta _{d}\left( 1-\eta
\right) \omega +\left( 1-\eta _{d}\right) \gamma },~~\text{for
large }\mu \text{. }  \label{IAB}
\end{equation}

Eve's Holevo function is $\chi _{EB}=S_{T}-S_{C}$, where $S_{T}$
is the von Neumann entropy of $\rho _{eE^{\prime }}$, while
$S_{C}$ is that of the conditional state $\rho _{eE^{\prime }|B}$.
From Eq.~(\ref{V-OUT}) consider
the block%
\begin{equation}
\mathbf{V}_{eE^{\prime }+}=\left(
\begin{array}{cc}
\mathbf{V}_{eE^{\prime }} & \mathbf{C} \\
\mathbf{C}^{T} & \mathbf{V}_{+}%
\end{array}%
\right) ,  \label{VEVE-BOB}
\end{equation}%
where%
\begin{equation}
\mathbf{V}_{eE^{\prime }}=\left(
\begin{array}{cc}
\omega \mathbf{I} & \sqrt{\eta \left( \omega ^{2}-1\right) }\mathbf{Z} \\
\sqrt{\eta \left( \omega ^{2}-1\right) }\mathbf{Z} & \left[ \eta
\omega
+(1-\eta )\mu \right] \mathbf{I}%
\end{array}%
\right) ,~~~\mathbf{C=}\sqrt{\eta _{d}(1-\eta )}\left(
\begin{array}{c}
\sqrt{ \omega ^{2}-1 }\mathbf{Z} \\
\sqrt{\eta }(\omega -\mu )\mathbf{I}%
\end{array}%
\right) ~.  \label{VEVE-TOT}
\end{equation}
We compute the symplectic spectrum of $\mathbf{V}_{eE^{\prime }}$\
which is
given by $\left\{ \nu _{1},\nu _{2}\right\} \overset{\mu }{\rightarrow }%
\left\{ \omega ,(1-\eta )\mu \right\} $. As a consequence, the von
Neumann entropy $S_{T}=h(\nu _{1})+h(\nu _{2})$ takes the
asymptotic form
\begin{equation}
S_{T}\overset{\mu }{\rightarrow }h(\omega )+\log
_{2}\frac{e}{2}(1-\eta )\mu .  \label{STOT-EVE-ASY}
\end{equation}%
To compute the conditional term $S_{C}$, we note that, after the
homodyne detection of quadrature $q$ ($p$), Eve's CM is projected
onto the conditional form%
\begin{equation}
\mathbf{V}_{eE^{\prime }|B }=\mathbf{V}_{eE^{\prime
}}-\mathbf{C}\left( \Pi \mathbf{V}_{+}\Pi \right)
^{-1}\mathbf{C}^{T},  \label{VEVE-COND}
\end{equation}%
with $\Pi =$diag(1,0) ($\Pi =$diag(0,1)). From the CM of Eq.~(\ref{VEVE-COND}%
), after some algebra and taking the asymptotic limit, we obtain
the
following symplectic spectrum%
\begin{align}
& \bar{\nu}_{1}\overset{\mu }{\rightarrow }\sqrt{\frac{\left(
1-\eta \right) \left( \eta _{d}\omega +\left( 1-\eta \right)
\left( 1-\eta _{d}\right)
\gamma \right) }{\eta \eta _{d}}\mu ,}  \label{NIC1} \\
& \bar{\nu}_{2}\overset{\mu }{\rightarrow }\sqrt{\frac{\omega
\left[ \eta _{d}+(1-\eta )\left( 1-\eta _{d}\right) \omega \gamma
\right] }{\eta _{d}\omega +(1-\eta )\left( 1-\eta _{d}\right)
\gamma }},  \label{NIC2}
\end{align}%
which provides the conditional entropy $S_{C}=h(\bar{\nu}_{1})+h(\bar{\nu}%
_{2})$.\ Combining this with Eq.~(\ref{STOT-EVE-ASY}), we derive
the asymptotic
Holevo bound%
\begin{equation}
\chi_{EB} \overset{\mu }{\rightarrow }h(\omega )-h\left( \bar{\nu}_{2}\right) +%
\frac{1}{2}\log _{2}\frac{\left( 1-\eta \right) \eta \eta _{d}\mu
}{\eta _{d}\omega +\left( 1-\eta _{d}\right) \left( 1-\eta \right)
\gamma }. \label{CHI}
\end{equation}%
Finally, using Eqs.~(\ref{IAB}) and (\ref{CHI}), we find formula
of the asymptotic key rate, which is given in
Eq.~(\ref{rate}).~$\blacksquare $



\begin{figure}[th]
\begin{center}
\begin{tabular}{c}
\includegraphics[height=7.5cm]{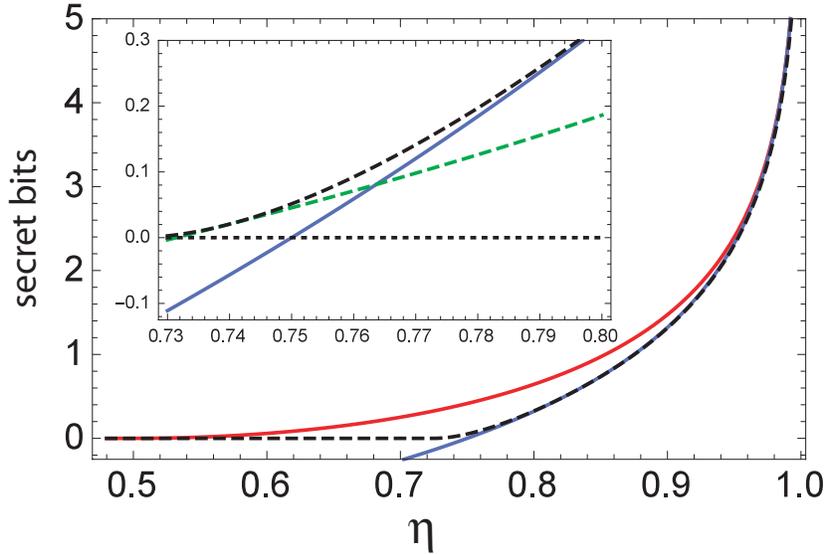}%
\end{tabular}%
\end{center}
\caption[example]{Secret bits versus transmissivity $\eta$ of a
thermal-loss channel with $\bar{n}_{th}=1$ mean photon (i.e.,
$\omega=3$). The main plot compares the rate $R_{M}(\eta ,\omega)$
of Eq.~(\ref{RM}) (black-dashed) with the upper bound (red line)
and lower bound (blue) of Eq.~(\ref{LBUB}). In the inset we zoom a
region of the main plot, where the optimal rate (black-dashed) is
strictly greater than the lower bound (blue) of Eq.~(\ref{LBUB}).
We also show the performance of the protocol in the case where Bob
employs a balanced beam splitter $\eta=1/2$ in a vacuum $\gamma=1$
(green dashed), which corresponds to Eq.~(4) of
Ref.~\citen{reversePRL}.} \label{MAIN}
\end{figure}

The secret key rate of Eq.~(\ref{rate}) depends on Bob's detection
parameters, i.e., transmissivity $\eta _{d}$ and variance $\gamma
$ (besides the parameters $\eta $ and $\omega $ of the channel).
It can therefore be maximized over $\eta _{d}\in \lbrack 0,1]$ and
$\gamma \geq 1$, providing the optimized rate $R_{M}(\eta ,\omega
)$ of Eq.~(\ref{RM}) which is our improved lower bound. It is easy
to check that $R_{M}(\eta ,\omega )$ outperforms all
previously-known lower bounds. It is sufficient to show that
the previous bounds can be retrieved for specific choices of the parameters $%
\eta _{d}$ and $\gamma $ in Eq.~(\ref{rate}). By setting $\eta
_{d}=1$ (no trusted noise at Bob's side), it is easy to verify
that we get $R(\eta ,\omega ,1,\gamma )=-\log _{2}\left( 1-\eta
\right) -h(\omega )$, corresponding to the lower bound of
Eq.~(\ref{LBUB}). According to the present derivation, a QKD
protocol achieving $-\log _{2}\left( 1-\eta \right) -h(\omega )$\
corresponds to Alice distributing TMSV states through the channel,
with Bob homodyning the output in $q$ or $p$; then, after many
rounds, Bob informs Alice of his choices, so that she
correspondingly homodynes the modes in her quantum memory to infer
Bob's outcomes. Then, by setting $\eta _{d}=1/2$ (balanced beam
splitter)\ and $\gamma =1$ (vacuum noise), we also find
\begin{equation*}
R(\eta ,\omega ,1/2,1)=\frac{1}{2}\log _{2}\frac{\omega +1-\eta
}{\left(
1-\eta \right) \left[ (1-\eta )\omega +1\right] }+h\left( \sqrt{\frac{\omega %
\left[ 1+(1-\eta )\omega \right] }{\omega +(1-\eta )\gamma
}}\right) -h(\omega ),
\end{equation*}%
which is Eq.~(4) of Ref.~\citen{reversePRL}. In Fig.~\ref{MAIN},
we numerically compare the maximized key rate $R_{M}(\eta
,\omega)$ of Eq.~(\ref{RM}) with respect to the other bounds,
considering a thermal-loss channel with $\bar{n}_{th}=1$ thermal
photon.


\section{Conclusions}\label{CONCLUSIONS}

In conclusion we have derived an improved lower bound for the
secret key capacity of the thermal-loss channel, by optimizing the
asymptotic key rate of a noise- and memory-assisted Gaussian QKD
protocol in reverse reconciliation. There is a strict separation
with respect to the previous best-known lower bounds derived in
Ref.~\citen{reversePRL}. Unfortunately, the improvement is small
and the gap with the upper bound is still open. Further efforts
must be devoted to close this gap and finally find the secret key
capacity of this very important bosonic channel.

\acknowledgments 

We acknowledge support from the EPSRC via the `Quantum
Communications HUB' (EP/M013472/1).

\end{document}